# Is Intelligence the Right Direction in New OS Scheduling for Multiple Resources in Cloud Environments?

XINGLEI DOU, LEI LIU*, LIMIN XIAO, Beihang University, China

**ABSTACT**

Making it intelligent is a promising way in System/OS design. This paper proposes OSML+, a new ML-based resource scheduling mechanism for co-located cloud services. OSML+ intelligently schedules the cache and main memory bandwidth resources at the memory hierarchy and the computing core resources simultaneously. OSML+ uses a multi-model collaborative learning approach during its scheduling and thus can handle complicated cases, e.g., avoiding resource cliffs, sharing resources among applications, enabling different scheduling policies for applications with different priorities, etc. OSML+ can converge faster using ML models than previous studies. Moreover, OSML+ can automatically learn on the fly and handle dynamically changing workloads accordingly. Using transfer learning technologies, we show our design can work well across various cloud servers, including the latest off-the-shelf large-scale servers. Our experimental results show that OSML+ supports higher loads and meets QoS targets with lower overheads than previous studies.

## 1 INTRODUCTION

The operating system (OS) is the most complex system software, often located at the core of almost every computer system. OSes manage the hardware, allocate the hardware resources, and execute the applications. The design of OSes affects the overall system performance and the user experiences [3,10,15,17,45,48,49,63].

Nowadays, OS can have a much heavier burden than previous days due to several reasons. (i) The hardware resources and the hierarchy become increasingly complicated, so managing diverse hardware and allocating resources accordingly in OS can be more challenging than before. (ii) The jobs that run on OS can be more diverse, e.g., latency-critical services in the cloud, interactive jobs, etc. And, as the number of computing units increases, many jobs can run concurrently. Therefore, scheduling jobs to meet user demands in a timely manner can be challenging. (iii) Modern OSes often have many tunable parameters that can affect performance, e.g., more than 17,000 kernel configurations in Linux kernel 5.1.0 [17,48]. Tuning these parameters can be time-consuming and ad-hoc. For example, even for the Linux memory system component, the parameter configurations for huge/small pages can significantly affect the overall system performance [57,58,59]. Thus, enabling these parameter configurations is challenging and essential to have an ideal performance.

We think that making OS intelligent can be a way to conquer the issues mentioned above. An intelligent OS kernel can learn from the environments by itself and, therefore, make appropriate decisions on how to allocate/schedule hardware resources to satisfy user demands. Even for those diverse applications that exhibit various behaviors and resource requirements, an intelligent OS can learn their patterns on the fly and enable the appropriate scheduling and configurations accordingly. In a word, with the learning ability, no matter how complicated the environments are, the OSes can handle them in a timely fashion.

Nevertheless, making OSes intelligent is still a long way to go. Many recent studies are proposed to enhance the key components in OS, e.g., memory management [3,60,62], job scheduling [13,36,44,53,61], storage management [2,4,43],

---

etc. In this work, to further show how an intelligent OS performs over traditional rule-/heuristic-based approaches and to show how much performance the intelligent mechanism can achieve, we design ML (machine learning)-based OS resource scheduling mechanisms OSML/OSML+ for scheduling co-located cloud services on large-scale servers. Our approaches use multiple collaborative ML models to predict QoS, shepherd the scheduling, and recover from QoS violations. We use complicated cloud environments in this study, showing the promise of intelligent OS. We make the following contributions in this study.

(1) We design OSML/OSML+ in our practice, showing that using ML models significantly improves scheduling exploration efficiency for multiple co-located cloud services and can handle the complicated resource sharing, under/over-provision cases timely. We deploy our work in reality and obtain the training data set and experiences for using intelligent systems.

(2) Our experiences show that using ML technologies to make OS intelligent is not that easy. We further show the underlying reason behind this, including model building, generalization, training, limitations and how to use models in an ideal way in OS design. Our design on OSML/OSML+ shows examples of these critical processes.

(3) We realize that instead of using a specific large model to handle all issues in OS scheduling, using multiple finer-grained ML models, each tackling one issue in scheduling, can achieve more accurate results. We propose a new multi-model collaborative learning framework for scheduling resources in the memory hierarchy, which is flexible to different tasks and generalizable to diverse platforms/applications. Our design in this paper shows that the multi-model collaborative approach has its respective advantages and can be a principle for ML-based OS design.

The following contents are organized as below. Section 2 shows some representative studies on leveraging ML technologies to make OS intelligent and further analyzes where to go if we want to make things better. Section 3 and 4 show the design details of OSML+ and the evaluations, through which we can see the technical details on how to make the OS resource scheduler intelligent. Section 5 summarizes how to leverage ML technologies to make OS intelligent, and section 6 makes the conclusion.

## 2  BACKGROUND AND MOTIVATIONS

Many studies try to make OS intelligent using ML technologies. These efforts cover a wide range of OS functions, e.g., resource scheduling for cloud services [1,3,5-9,11-12,14,19-20], I/O scheduling [2,51], load balancing and task scheduling [13,54,60], memory page allocation and reclamation [52,60,62], and parameter tuning for OS/system software [4,10,16]. We summarize some representative ones among these studies and show them in Table 1 in detail. Generally, these studies fall into three categories. For each category of studies, we analyze how they perform to make OS intelligent and how to conquer the shortcomings when using them in OS design as follows.

**The first category is the offline learning approach.** Offline learning often refers to ML models trained on a pre-collected data set. Offline learning includes statistical learning models (e.g., support vector machines in [1], boosted tree in [8]) and deep neural networks (e.g., multi-layer perceptron (MLP) in [3], convolutional neural networks (CNN) in [8]). Offline learning models can have low inference overhead and high accuracy. They work well with low overheads in environments that do not change drastically. For instance, LinnOS [2] in Table 1 employs a light neural network (NN) to infer per-I/O performance on SSD. Each inference only takes 4-6us, less than the access latency of current SSDs (e.g., 80us). Therefore, the model can be used frequently for fine-grained tasks. However, what challenges the offline learning approaches is the generalization. Some studies make offline models generalizable by using training samples covering a broader range of the state space. For example, OSML [3] in Table 1 allocates core/cache resources for LC services using offline models trained based on a large data set covering widely used cloud services, making OSML generalizable across services with diverse computing and memory patterns. In terms of generalization for diverse platforms, OSML needs transfer learning and more training data. Moreover, some other studies reduce the state space by selecting critical features. For example, FIRM [1] in Table 1 uses an SVM with a small state space to identify microservices that cause QoS violations. The model has a small state space with only two input features. The two features are only related to response latency and are platform-independent, making the model generalizable across platforms. *Regarding the new OS* *design, offline learning approaches work well in cases where the environment does not have drastic changes. It also requires sufficient training data for generalization.*

Table 1. Representative studies on using ML technologies to make system intelligent

| Studies | ML Models | Function | Goals | Generalization | Pros | NI |
|---|---|---|---|---|---|---|
| FIRM [1] | Offline classification (SVM) | Identifying microservices that cause QoS violations | Filtering microservices that need reprovision | Generalizable due to limited state space | Simple and efficient | Accuracy can be improved |
| | Reinforcement learning (DDPG) | Scheduling CPU time limit, memory/IO/ network bandwidth, LLC ways for microservices | Mitigating QoS violations | Generalizable via online learning | Preventing QoS violations | High retraining overheads |
| LinnOS [2] | Offline classification (NN) | Identifying slow I/O on SSD | Revoking slow I/O for higher I/O performance | Amplified errors for unseen cases | Low inference overhead | Generalization can be improved |
| OSML [3]/ OSML+ | Offline MLP | Predicting an LC service's CPU cores/ LLC ways/memory bandwidth requirements | Achieving near-optimal resource allocation | Generalizable due to extensive training data | Predicting near-optimal solution | Training data set is large |
| | Offline MLP | Predicting QoS slowdown of an LC service after depriving resources | Trading QoS for resources | Generalizable due to extensive training data | Preventing QoS violations | Training data set is large |
| | RL (DQN/DDPG) | Scheduling CPU cores/ LLC ways for LC services | Mitigating QoS violations | Generalizable due to online learning | Handling over-/under- provision | Retraining overheads |
| iBTune [4] | Offline classification (DNN) | Predicting upper bounds of the request response times after shrinking cloud databases' buffer sizes | Saving memory while guaranteeing QoS/SLA | Amplified errors for unseen cases | Pairwise DNN for better generalization | Generalization can be improved |
| Cilantro [5] | Reward-based policy (UCB) | Scheduling CPU cores for microservices | Minimizing latency, improving performance | Generalizable due to online learning | Low overhead, using no prior knowledge | Needs resampling when workloads change |
| AWARE [6] | Reinforcement learning (DDPG) | Scheduling CPU time limit/memory limit for containers | Improving CPU/memory utilization, protecting QoS | Generalizable due to online learning | Efficient RL model training | Generalization can be improved |
| Twig [7] | Reinforcement learning (BDQ) | DVFS and CPU core scheduling for LC services | Reducing energy usage, protecting QoS | Generalizable due to online learning | Coordinating multiple learning agents | High retraining overheads |
| Sinan [8] | Offline regression (CNN) | Predicting response latency of the next timestep | Generating intermediate results for predicting QoS | Robust to workload changes | Explainable ML model | Generalization can be improved |
| | Offline classification (Boosted Tree) | Predicting whether a QoS violation would occur | Searching the lowest CPU time limit that can meet QoS | Robust to workload changes | Low overhead | Generalization can be improved |
| CLITE [9] | Reward-based policy (Bayesian Optimization) | Scheduling CPU cores, LLC ways, memory bandwidth for cloud services | Protecting QoS, improving BE throughput | Generalizable due to online learning | Low overhead, using no prior knowledge | Needs resampling when workloads changes |
| PSS [10] | Reinforcement learning (Online perceptron) | Providing a prediction service for different optimization problems | Eliding hardware lock/tuning JIT parameter/guiding page reclamation for application speedup | Generalizable due to online learning | A system service for general optimization goals | Accuracy can be improved |

**The second category is the reinforcement learning (RL).** RL models learn to maximize a reward by interacting with environments. Deep Deterministic Policy Gradient (DDPG) used in [1,6], Deep Q-network (DQN) used in [3] are representative RL algorithms. They can learn online from historical scheduling decisions and learn from the feedback from the environment, making them resilient to changes in the environment and the workloads. RL works well for long-term scheduling tasks where the workloads and environment may change. For example, the work in [10] in Table 1 builds a prediction system service using an online perceptron, achieving application speedup by guiding hardware lock elision, tuning parameters for PyPy's Just-In-Time (JIT) compiler, and guiding page reclamation. The effort in [56] uses RL to avert VM failures in the cloud environment. The study in [21] guides data placement in hybrid storage systems using Q-learning. Moreover, several studies [3,6,7,18,19] leverage RL to schedule interactive resources (e.g., CPU cores, LLC ways, memory/IO/network bandwidth) for co-located cloud services, protecting QoS of latency-critical services and improving the performance for throughput-oriented services. Resource scheduling for co-located applications is a notorious complicated issue, as the co-located cloud services exhibit fluctuating loads and drastically different behaviors across the memory/storage hierarchy. RL is an ideal approach. The studies in [1,6] in Table 1 further reduce the training overheads for RL models using transfer learning or bootstrapping. *To sum up, for the new OS design, RL approaches can handle the cases in which the environments always change, and can also work well to handle the complicated interactive cases where the OSes need to have solutions in large exploration spaces.*

**The third category is the reward-based algorithms.** Reward-based algorithms build a model that learns from sampling points and corresponding rewards. They infer the optimal solution using the model. They do not need offline traces and can achieve a near-optimal solution faster via sampling in the scheduling exploration space and observing the real-time feedback. For example, Bayesian optimization models the reward as a probability distribution and uses this model to guide the search behaviors (e.g., the study [9] in Table 1). The Upper Confidence Bound (UCB) strategy in the context of multi-armed bandit problems assigns each option a UCB based on past observations and selects the option with the highest upper confidence bound, e.g., the work [5] in Table 1. They use reward-based algorithms to find the optimal resource partition among co-located cloud services. Reward-based algorithms repeat sampling in the scheduling exploration space and use the sampling points to update the model. The allocation policy that yields the highest reward will be used. They can find a feasible resource allocation within tens of sampling steps. *Regarding the new OS design, the reward-based policies require no prior knowledge and can converge fast. Yet, they should be used carefully, as they incur performance fluctuations during sampling and require resampling when the workloads or the environment change.*

**Is ML a silver bullet for new OS design?** Through Table 1, we can see that although ML technologies bring benefits for OS design, most of the studies face the problem of generalization (e.g., retraining overheads, sensitive to new platforms, etc.), accuracy and performance in unseen cases. Moreover, how to select/build an appropriate ML model accordingly? Is a global model sufficient to handle all cases? Or, to use per-task/application models separately?

To study these problems, we propose OSML+ (a new design based on OSML [3]), a typical instance of using ML in OS design. OSML+ is an intelligent resource scheduling mechanism for co-located cloud services. Model building in OSML+ uses multiple models (static MLPs plus dynamic DDPG) instead of a large global model; each model has its own duty. So, OSML's models are lightweight, their functions are clearly defined, and it is easy to locate and debug the problems. OSML has an extensive data set used for training. Thus, the models can be trained and generalized on new platforms. Finally, we propose a new framework in OSML+ for orchestrating ML models to accomplish the scheduling goals. The new framework makes the OSML+ easier to generalize.

Through our series studies on OSML+/OSML [3], we want to show examples of making OS intelligent, including building ML models, designing the supporting framework, and making it generalized.

## 3 THE DESIGN OF OSML+
### 3.1 Resource Scheduling for Co-located Cloud Services
For cost-efficiency, multiple services are usually co-located on a server, including latency-critical (LC) services with strict QoS targets and throughput-oriented best-effort (BE) services [3,9,19,23]. Runtime resource scheduling is the core for quality of service (QoS) control in these complicated co-location cases [3,9,23,25]. Resource schedulers schedule interactive resources, e.g., CPU cores, cache, and memory bandwidth, reducing resource contentions among co-located services and protecting the QoS. How to design new resource scheduling mechanisms is a challenging job in this era [3,9,23,26,32]. In this section, we identify several key challenges that make resource scheduling for co-located cloud services complicated and time-consuming. We further illustrate the difficulties of addressing these challenges using prior schedulers (e.g., heuristic-based approaches) and show that leveraging ML can be a promising solution.

**Challenge 1 - Achieving the optimal solution efficiently in the large resource scheduling exploration space.** Today's data center servers have an increased number of CPU cores, larger LLC capacity, larger main memory capacity, higher bandwidth, and the resource scheduling exploration space becomes much larger than ever before as a result [3,9,23,26]. The large resource scheduling exploration space, which consists of diverse resources, prohibits the schedulers from achieving the optimal solution quickly [3,9,23,24,26]. Additionally, co-located cloud services exhibit various behaviors across the memory/storage hierarchy. They share multiple interactive resources such as CPU cores, cache, memory/IO bandwidth, and main memory banks, leading to resource contentions among co-located services. These issues make resource scheduling for co-located cloud services difficult and time-consuming.

Previous heuristic resource schedulers [23,26] explore the scheduling space in a fine-grained way, i.e., adjusting one resource dimension at a time until the QoS is met. This approach takes tens of seconds or even minutes to find a resource scheduling solution in practice [23,26]. *By contrast, using offline ML models can efficiently achieve the Optimal Allocation Area (**OAA**) [3] for LC services in the large scheduling exploration space, which satisfies the QoS targets using appropriate resources (in Figure 1). This is because offline ML models learn the relationship between the scheduling solutions and the QoS demands from historical traces in the data set. As a result, ML models can predict an optimal allocation (or near optimal) without sampling in the scheduling exploration space in a fine-grain way.*

**Challenge 2 - Avoiding QoS fluctuations incurred by Resource Cliff.** We observe the Resource Cliff (**RCliff**) phenomenon commonly exists in many widely used cloud services, in which a slight resource reduction could incur a significant performance slowdown for LC services. RCliff is defined as the resource allocation cases that could incur the most significant performance slowdown if resources (e.g., core, cache) are deprived via a fine-grain way in the scheduling exploration space. RCliff can hardly be avoided for existing resource schedulers; therefore, we can observe the QoS always fluctuates severely. This is because these schedulers need to sample various allocations during the scheduling process, inevitably incurring a sharp QoS slowdown if falling off the RCliff [3,23,26].

We study all LC services listed in Table 2 and find that the RCliff phenomenon always exists for LC services. We show several of them in Figure 1. We show how the response latency of LC services changes with variations in the allocation of critical resources, i.e., CPU cores and LLC ways. The red line highlights the RCliff in Figure 1 (can be obtained by using the knee solution [46]). As shown in Figure 1-a, in the cases where 6 cores are allocated to Moses, the response latency increases significantly from 34ms to 4644ms if merely one last level cache (LLC) way is reduced (i.e., from 10 ways to 9 ways). Similar phenomena also happen in cases where computing resources are reduced. In Figure 1, We also highlight each LC service's OAA in the scheduling exploration space, i.e., the ideal number of allocated cores and LLC ways to bring an acceptable QoS. More resources than OAA cannot deliver more significant performance, but fewer resources lead to the danger of falling off the RCliff. OAA is the goal that schedulers should achieve.

**Scheduling.** Existing resource schedulers have shortcomings in finding OAAs and avoiding RCliffs effectively and efficiently. Many schedulers often employ heuristic scheduling algorithms [23,26], i.e., they increase/reduce resources until the monitor alerts that the system performance is suffering a severe slowdown. Yet, these approaches could incur unpredictable latency spiking. For example, if the current resource allocation for an LC service is in the base of RCliff (i.e., the yellow color area in Figure 1), the scheduler has to try to achieve OAA. However, as the scheduler doesn't know the "location" of OAA in the scheduling exploration space, it has to increase resources step by step in a fine-grain way. Thus, the entire scheduling process from the base of the RCliff will incur very high response latency. For another example, if the current resource allocation is on the RCliff or close to RCliff, a slight resource reduction for any purpose could incur a sudden and sharp performance drop for LC services. The previous efforts [3,23,26] find there would be about hundreds/thousands of times latency jitter, indicating the QoS cannot be guaranteed during these periods. Thus, RCliffs should not be neglected when designing a scheduler. *Using offline ML models can intelligently avoid RCliffs. For instance, offline training using the data in Figure 1 can make the model learn the knowledge of where the RCliffs are in the scheduling spaces. Thus, the ML-based schedulers can directly avoid allocating the resources around RCliffs without using the try-and-error approaches.*

**Challenge 3 - Handling dynamic resource demands and scheduling interactive/fungible resources.** In practice, cloud services exhibit dynamic changes in loads and required resources. Resource schedulers need to adjust the resource allocations online in a timely fashion. If the applications' behaviors change, the schedulers must immediately provide new resource allocation solutions in the scheduling space (Figure 1). Otherwise, it will lead to resource over-provision or under-provision, negatively affecting the QoS or wasting resources. For more complicated cases where several cloud services are co-located on a specific server, adjusting one application's resource allocations may affect other applications' allocations. So, the scheduler must quickly have a global solution for all co-located applications. This is a challenging job. *Using reinforcement learning (RL) is a practical approach to meet dynamically changing resource demands in such complicated cases. It is adaptive to changes as it continuously learns from real-time feedback, making it*

Table 2. Latency-critical (LC) services [26,47]. The max load is the highest Requests Per Second (RPS) an LC service can reach while satisfying the QoS. The max load of an LC service may vary across platforms.

| ID | LC service | Domain | RPS (Requests Per Second) |
|----|-----------|--------|---------------------------|
| 1 | Img-dnn [37] | Image recognition | 2000,3000,4000,5000,6000 (Max) |
| 2 | Masstree [37] | Key-value store | 3000,3400,3800,4200,4600 |
| 3 | Memcached [39] | Key-value store | 256k,512k,768k,1024k,1280k |
| 4 | MongoDB [38] | Persistent database | 1000,3000,5000,7000,9000 |
| 5 | Moses [37] | RT translation | 2200,2400,2600,2800,3000 |
| 6 | Nginx [40] | Web server | 60k,120k,180k,240k,300k |
| 7 | Specjbb [37] | Java middleware | 7000,9000,11000,13000,15000 |
| 8 | Sphinx [37] | Speech recognition | 1,4,8,12,16 |
| 9 | Xapian [37] | Online search | 3600,4400,5200,6000,6800 |
| 10 | Login [47] | Login | 300,600,900,1200,1500 |
| 11 | Ads [47] | Online renting ads | 10,100,1000 |

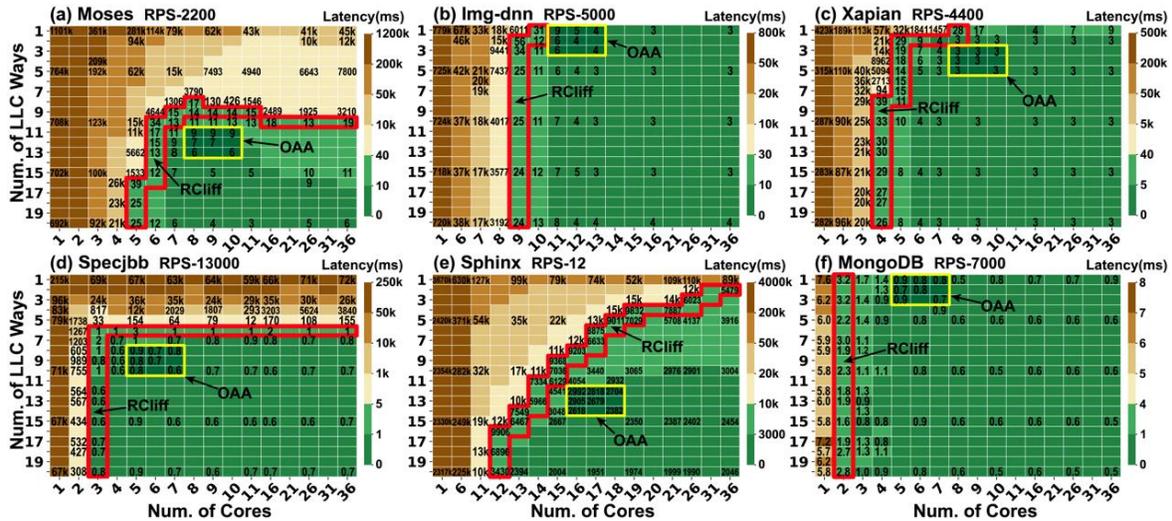

Figure 1. The resource scheduling exploration space for CPU cores and LLC ways. All LC services here are with 36 threads. Each column/row represents a specific number of CPU cores/LLC ways allocated to an LC service. Each cell denotes the LC service's response latency under the given number of cores and LLC ways [3]. The experiments are conducted on server 1 in Table 4.

*suitable for resource scheduling in dynamic changing environments.*

Additionally, scheduling a combination of resources (e.g., cores, LLC ways) with complicated interactions is challenging. Prior studies [23,24,26] show that the CPU cores, cache hierarchy, and memory bandwidth are interactive resources. Solely considering a single dimension during the scheduling process often leads to sub-optimal QoS [3,9,23,24]. Existing schedulers using heuristic algorithms usually schedule one dimension resource at a time, bringing high overheads on scheduling multiple interactive resources [23,24]. *Regarding ML approaches, using an extensive training data set, ML models can learn the interactive correlation among various interactive resources in diverse cases. So they can quickly make the schedulers achieve the ideal allocations for multiple interactive resources.*

**Challenge 4 - Tackling the complexity of the scheduling process.** Resource scheduling is a complicated job, which usually has several independent sub-tasks. Using a global model to handle these sub-tasks is challenging, as each sub-task has its respective aims and optimization goals [1,3,8]. For example, the scheduler designed in [3] has three sub-tasks in its scheduling process (i.e., achieve OAA/avoid RCliff, predict QoS, shepherd allocations on the fly). And [3] has three well-designed lightweight ML models to meet the three sub-tasks separately. The ML models work collaboratively in a pipe-lined way to fulfill a specific scheduling job. Using the design of a multi-model collaborative approach, the scheduling function is decomposed, making locating and debugging the problems easy. Moreover, compared with a global model, lightweight-specific models can have higher accuracy and better generalization ability (e.g., the state space in the training period is reduced). In this mind, the critical problem will be the design of a framework that combines multiple ML models and makes them work smoothly. The framework works as the central logic of the schedulers. Without a well-designed framework, ML models cannot work together to handle complicated scheduling tasks.

### 3.2 Building ML Models for Intelligent Scheduling

To this end, we propose OSML+, a multi-model collaborative learning-based approach for intelligent resource scheduling. OSML+ has multiple ML models, each handling one of the challenges in the scheduling process mentioned above. (i) To achieve the optimal allocations and avoid RCliffs in the large scheduling exploration space efficiently, OSML+ employs an offline MLP model (Model-A) to predict the optimal allocation that satisfies the QoS using the minimum resources (i.e., OAA). (ii) To avoid try-and-error scheduling approaches for diverse resource demands, OSML+ uses an offline MLP (Model-B) to predict the QoS when a new resource allocation is conducted. (iii) To handle the dynamic changes in co-located services, OSML+ employs an RL agent (Model-C) to shepherd the allocations dynamically, handling QoS violations and resource over/under-provision cases online.

**(1) Building ML model for predicting the OAA and RCliff.** We denote this model as Model-A. We use an offline multi-layer perceptron (MLP) in Model-A, providing the (near) optimal allocations for applications. The neural network has three hidden layers with 32, 64, and 32 neurons, respectively. Each layer is a set of nonlinear functions of a weighted sum from all outputs fully connected from the prior one [27]. The increase from 32 to 64 neurons aims to capture more complex features. The decrease from 64 to 32 neurons aims to refine these features. This design makes the network understand the data better. There is a dropout layer with a loss rate of 30% behind each hidden layer to prevent overfitting. For each LC service, the **inputs** of Model-A are shown in Table 3. They include the performance traces (i.e., IDs 1-6 in Table 3), resource usage of the LC service (i.e., IDs 7-9), and resource usage of neighboring applications (i.e., IDs 10-12). The **outputs** of Model-A include the OAA for multiple interactive resources, OAA bandwidth (bandwidth requirement for OAA), and the RCliff. Using the MLP for predicting OAA and RCliff is practical. The reasons are multi-fold. (1) An offline ML model can be trained for a long period, giving the model sufficient time to learn the relationship among the QoS demands, resource allocations and architectural hints (e.g., IPC, cache misses, memory footprint, etc.) within the data set. (2) The data set can have diverse and representative data samples that cover a wide range of the state space, making the model generalizable for unseen cases.

**Model-A training**. Collecting training data is an offline job. On Server 1 in Table 4 [3], we collect the performance traces that involve the parameters in Table 3 for LC services in Table 2. The parameters are normalized into [0,1] according to the function: Normalized_Feature = (Feature-Min)/(Max-Min). Feature is the original value; Max and Min are predefined according to different metrics. For each LC service with every common RPS demand, we sweep 36 threads to 1 thread across LLC allocation policies ranging from 1 to 20 ways, map the threads on a certain number of cores and collect the performance traces accordingly. In each case, we label the corresponding OAA, RCliff and OAA bandwidth. For example, Figure 2 shows a data collection case where 8 threads are mapped onto 7 cores with 4 LLC ways. We feed the LC services with diverse RPS (Table 2), covering most of the common cases. Moreover, we map LC services on the remaining resources in the above process and get the traces for co-location cases. Note that the resources are partitioned among applications. Finally, we collect 43,299,135 samples (data tuples), covering 1,308,296 allocation cases with different numbers of cores, LLC ways, and bandwidth allocations. A large amount of traces may lead to higher accuracy for ML models. The workload characteristics are converted to traces consisting of hardware parameters used for fitting and training MLP to provide predictions.

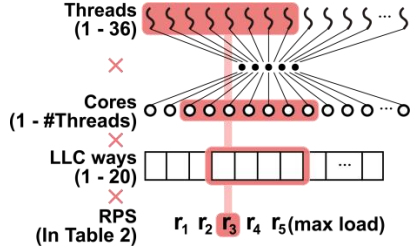

Figure 2. Model-A data collection

Table 3. The input parameters for ML models.

| ID | Feature | Description | Used in model | | |
|---|---|---|---|---|---|
| | | | Model-A | Model-B | Model-C |
| 1 | IPC | Instructions per clock | √ | √ | √ |
| 2 | Cache Misses | LLC misses per second | √ | √ | √ |
| 3 | MBL | Local memory bandwidth | √ | √ | √ |
| 4 | CPU Usage | The sum of each core's utilization | √ | √ | √ |
| 5 | Virt. Memory | Virtual memory in use by an app | √ | √ | |
| 6 | Res. Memory | Resident memory in use by an app | √ | √ | |
| 7 | Allocated Cores | The number of allocated cores | √ | √ | √ |
| 8 | Allocated Cache | The capacity of allocated cache | √ | √ | √ |
| 9 | Core Frequency | Core Frequency during run time | √ | √ | √ |
| 10 | Expected Cores | Expected cores if the allocation changes | | √ | |
| 11 | Expected Cache | Expected cache if the allocation changes | | √ | |
| 12 | Cores used by N. | Cores used by Neighbors | √ | √ | |
| 13 | Cache used by N. | Cache capacity used by Neighbors | √ | √ | |
| 14 | MBL used by N. | Memory BW used by Neighbors | √ | √ | |

**Applying transfer learning (TL) to new platforms.** Offline learning models require sufficient training data for generalization. Collecting extensive training data on each new platform is impractical due to high costs. We address this challenge using transfer learning. The data set collected on Server 1 has a rich set of information. ML models can be trained and generalized based on these data and then used on new platforms with low-overhead transfer learning [55]. On the two new platforms (i.e., Server 2 and Server 3 in Table 4), we collect traces only for hours. We only collect performance traces for cases where the number of launched threads equals the number of CPU cores on each new platform. For instance, we collect 422,400 data samples on Server 2 in Table 4 and 290,400 on Server 3 in Table 4. We copy the network parameters of Model-A trained on Server 1 to the models on the new platforms, and then train the models using data from the new platforms for transfer learning. Using this TL approach, the offline models can be generalized to new platforms with a low data collection overhead.

Table 4. The specifications of platforms used in this study.

| Configurations/ Servers | Server 1 | Server 2 | Server 3 |
|---|---|---|---|
| CPU (×2 Sockets) | Intel Xeon E5-2697 v4 | Intel Xeon Gold 6338 | Intel Xeon Gold 5220R |
| Logical Processor Cores | 36 Cores (18 phy. cores) | 64 Cores (32 phy. cores) | 48 Cores (24 phy. cores) |
| Processor Speed | 2.3GHz | 2.0GHz | 2.2GHz |
| Main Memory / Channel per-socket / BW per-socket | 128GB, 2400MHz DDR4 / 4 Channels, 76.8 GB/s | 256GB, 2933MHz DDR4 / 4 Channels, 94.0 GB/s | 128GB, 3200MHz DDR4 / 4 Channels, 102.4 GB/s |
| L1I, L1D & L2 Cache Size | 32KB, 32KB and 256KB | 32KB, 48KB and 1.25MB | 32KB, 32KB and 1MB |
| L3 Cache Size | 45MB - 20 ways | 48MB - 12 ways | 35.75MB - 11 ways |
| Disk | 1TB, 7200 RPM, HD | 2TB, SSD | 500GB, SSD |
| GPU | NVIDIA GP104 [GTX 1080], 8GB Memory | NVIDIA GeForce RTX 3080 LHR | NVIDIA GP104 [GTX 1080], 8GB Memory |

**(2) Building ML model for predicting the QoS.** We denote this model as Model-B. Model-B predicts the achieved QoS when a new resource allocation is conducted. It employs an MLP with the same structure in Model-A plus two more input items, i.e., the expected cores and expected cache in Table 3 when a new resource allocation is conducted. Model-B **outputs** the achieved QoS for the to-be-conducted allocation behaviors. Using an offline model for QoS prediction is practical, because the model learns from a large amount of typical traces, fitting the correlation between the QoS variations and diverse resource allocation changes.

Model-B can trade QoS for resources. It can predict whether the QoS is violated when depriving an LC service of a specific amount of allocated resources. For example, on a server without idle resources, when an LC service needs m more cores and n more LLC ways to meet its QoS target, Model-B can be used to predict how much QoS slowdown would the neighboring application suffer if it is deprived of a number of CPU cores and LLC ways. Finally, OSML+ finds the best solution to match <m, n>, which has a minimal impact on the co-located applications' current allocation state. Moreover, Model-B saves the time required for sampling in the scheduling exploration space. When having a new allocation policy in the scheduling exploration space, instead of conducting the allocation and observing the response latency changes, using Model-B can obtain the predicted feedback much faster. We use Model-B's predicted QoS as a component in Model-C's reward function for saving sampling overhead during Model-C's online training. Details are in the design of Model-C's reward function.

**Model-B training.** We build the data set for Model-B training based on traces collected for Model-A training. For each data tuple in these traces, we randomly generate the resource allocation when a resource scheduling action is conducted (i.e., expected cores and expected cache), and use the achieved QoS for the new allocation as the label for the data tuple. For Model-B, we have 43,299,135, 422,400, and 290,400 data tuples on Server 1, Server 2 and Server 3 in Table 4, respectively. The data set on Server 1 covers more cases than the data set on Server 2 and Server 3. We train Model-B using data from Server 1 [3]. Then, we employ TL to train Model-B on the two new platforms based on the model trained on Server 1, enhancing the model's generalization performance with minimal data collection overhead on new platforms. We use the same TL approach in Model-A training.

**(3) Building RL model for handling dynamic changes.** We denote this model as Model-C, which is a reinforcement learning model that dynamically shepherds the allocations and corrects the resource under-/over-provision. Model-C learns online to conduct resource scheduling actions to maximize long-term rewards. Figure 3 shows Model-C in a nutshell. Using an RL model to handle dynamic changes is a promising approach. As OSML+ is designed for co-located cloud services that run for a long period, these co-located services may exhibit varying loads and resource requirements over time. Model-C learns online and is resilient to changes in loads and required resources, constantly protecting the QoS for co-located cloud services.

Model-C's core component is a deep deterministic policy gradient (DDPG) model [28], consisting of four deep neural networks, i.e., Actor network, Critic network, Actor target network and Critic target network. Each network has three hidden layers with 32, 64, and 32 neurons, respectively. The Actor network is used to infer a resource scheduling action $\pi(S_t) = A_t$ each time it receives the system status $S_t$ (shown in Table 3). The Critic network infers the Q-value $Q(S_t, A_t)$, i.e., the expected cumulative reward of conducting the action $A_t$ given the state $S_t$. The Actor target and Critic target networks have identical structures as the Actor and Critic networks. They are updated using soft updating to avoid fluctuations in the target values, enhancing the stability of the online learning process. We use softmax as the activation function in the Actor network and Actor target network, thereby transforming resource scheduling into a classification problem. All other layers use ReLU as the activation function. Each time OSML+ schedules online using Model-C, it forwards the current system status $S_t$ to the Actor network to obtain the resource scheduling action $A_t$. OSML+ conducts the action and calculates the reward $R_t$. Then, the system status transitions to $S_t$'. The tuple $<S_t, A_t, R_t, S_t'>$ is saved in the experience pool for experience replay. The model injects Gaussian noise with mean $\mu$ and standard deviation $\sigma$ into the Actor network's output to promote exploration in the learning process. This approach prevents Model-C from getting stuck in local optima.

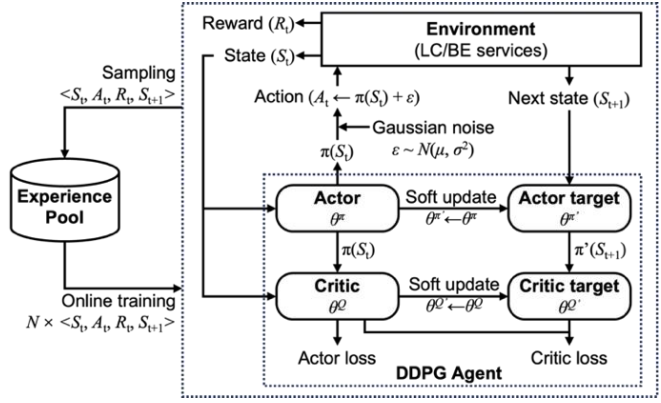

Figure 3. Model-C in a nutshell.

Table 5. The action space of Model-C.

| Action ID | Scaling Direction | Resource |
|---|---|---|
| 0 | Scale up | CPU cores |
| 1 | Scale down | |
| 2 | Scale up | LLC ways |
| 3 | Scale down | |
| 4 | Scale up | memory bandwidth |
| 5 | Scale down | |
| 6 | Idle | |

**Model-C's action space.** Model-C's action space contains seven resource scheduling actions that control three types of resources, i.e., CPU cores, LLC ways, and memory bandwidth, shown in Table 5. Each resource type has an action for scaling up the allocation for an LC service and another for scaling down the allocation. For actions related to the three types of resources, the step size is 1 CPU core, 1 LLC way, and 10% memory bandwidth, respectively. Additionally, there is an idle action that does not adjust any resources. Model-C predicts a resource scheduling action in the action space each time it receives the system status. The Actor network outputs a vector $A = \{a_1, a_2, ..., a_k\}$, $\sum a_i = 1$, where $a_i$ represents the probability of the $i$-th action in the action space. The action with the highest probability is executed.

**Model-C's reward.** The training of the reinforcement learning model's online learning incurs high costs as it needs to wait and observe the changes in response latency after conducting a resource scheduling action. Model-C avoids this cost by using Model-B's predicted QoS as a reward component, achieving faster model convergence. The details of Model-C's reward function design are as follows. OSML+ has two optimization goals, i.e., protecting the QoS of LC services and improving the throughput of BE (best-effort) services. The reward function is designed as a piecewise function. If the QoS of any LC service is not met, OSML+ optimizes the QoS; if the QoS of all LC services are met, OSML+ reclaim over-provisioned resources for higher BE throughput. The reward function is shown as follows:

$$r = \begin{cases} r_{QoS} & \text{if any LC service's QoS is not met} \\ 2 + r_{resource} & \text{if all LC services' QoS is met} \end{cases}$$

$$r_{QoS} = I(\text{predicted\_QoS}) + \frac{1}{N} \sum_{i}^{N} \min(1, \frac{\text{QoS\_target}_i}{\text{Latency}_i})$$

$$r_{resource} = \frac{1}{M} \sum_{i}^{M} (1 - \frac{\text{Resource\_usage}_i}{\text{Resource\_limit}_i})$$

OSML+ only optimizes the QoS when the QoS of any LC service is not met, i.e., $r = r_{QoS}$. The QoS reward $r_{QoS}$ is designed to make the LC services recover from QoS violations. The better the QoS, the higher the QoS reward value. OSML+ learns to adjust resources to recover from QoS violations, aiming to achieve a higher QoS reward. The QoS reward has two components: (1) $I(\text{predicted\_QoS})$, which indicates whether the QoS predicted by Model-B is satisfied. Model-B outputs the predicted_QoS for the new resource allocation. When the predicted QoS meets the QoS target, $I(\text{predicted\_QoS}) = 1$. Otherwise, $I(\text{predicted\_QoS}) = 0$. This component is designed to provide OSML+'s Model-C with instant feedback. With this component, OSML+'s Model-C does not have to wait for response latency changes after conducting a resource scheduling action. Thus, Model-C can achieve a shorter scheduling time interval, reducing the convergence time. Moreover, this component has a high accuracy of 96.4%. This indicates that using the predicted QoS in reward is practical and reliable. (2) $1/N\sum_i^N \min(1, \text{QoS\_target}_i/\text{Latency}_i)$, which indicates the average QoS of all LC services on-the-fly. $N$ is the number of LC services on-the-fly. This component returns the highest reward value of 1 when all LC services meet their QoS targets, and it returns a reward of less than 1 when QoS targets can not be met. With this component, OSML+ learns to recover from QoS violations by adjusting the resource allocations. $r_{QoS}$ yields a reward value between 0 and 2. A higher value of $r_{QoS}$ indicates that the resource scheduling action brings better QoS for LC services running on the server. OSML+ reduces the resource usage for LC services when all LC services meet their QoS targets, i.e., $r = 2 + r_{resource}$. The constant value 2 is the maximum value of $r_{QoS}$. This results in a reward value higher than 2 when all LC services' QoS is met. The term $r_{resource}$ represents the average percentage of resources that BE services can use. It yields a higher reward value when the LC services use fewer resources. In $r_{resource}$, $M$ represents the number of resource types under control. Resource_usage represents the resources occupied by all LC services. Resource_limit represents all available resources on the platform.

**Online learning.** Model-C collects online traces. The online learning process of Model-C is illustrated in Figure 3. Each time OSML+ schedules using Model-C, a batch of tuples is extracted from the experience pool to update Model-C's networks. The Mean Squared Error (MSE) between the predicted Q-value and the target Q-value is used as the loss function for the Critic network. Specifically, the predicted Q-value is calculated as $Q(S_t, A_t)$, where Q represents the action-value function of the Critic network. The target Q-value is calculated as $y_t = R_t + \gamma Q'(S_{t+1}, \pi'(S_{t+1}))$. Here, $Q'$ represents the action-value function of the Critic network; $\pi'$ represents the policy function of the Actor target network; $\gamma$ ($\gamma \in [0, 1]$) is a discount factor that defines the weight of immediate rewards and future rewards, default as 0.99. The loss function for the Actor network is defined as the negative Q-value, aiming to maximize the expected cumulative reward. The networks are updated by minimizing the loss using gradient descent. The Critic target and the Actor target networks are updated using soft updating, i.e., $\theta^Q = \tau\theta^Q + (1 - \tau)\theta^Q$, $\theta^{\pi'} = \tau\theta^{\pi'} + (1 - \tau)\tau\theta^{\pi'}$. Here, $\theta$ represents the network parameters. $\theta^Q$, $\theta^Q$, $\theta^\pi$, $\theta^{\pi'}$ denotes the parameters in the Critic, Critic target, Actor, and Actor target networks, respectively. The hyperparameter $\tau$ ($\tau \in [0, 1]$) controls the update rate of the target networks, default as 0.001. The target network's parameters are updated using soft updating, where the parameters of the Actor and the Critic networks are blended into the Actor target and Critic target network. Doing so improves the stability of the training and prediction.

Compared with the prior work OSML [3], OSML+'s Model-C converges faster by using a shorter scheduling time interval. OSML schedules every 1 second because it has to wait and observe the response latency changes to obtain the feedback for a resource scheduling action [3]. By contrast, OSML+ uses Model-B's predicted QoS as a reward component in Model-C and does not have to wait for the response latency changes. Therefore, OSML+ can conduct more scheduling actions during the same scheduling phase for faster convergence.

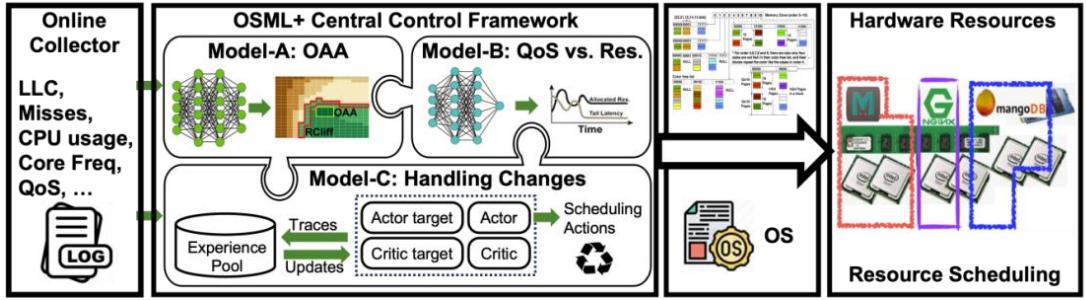

Figure 4. The overview of OSML+'s system architecture.

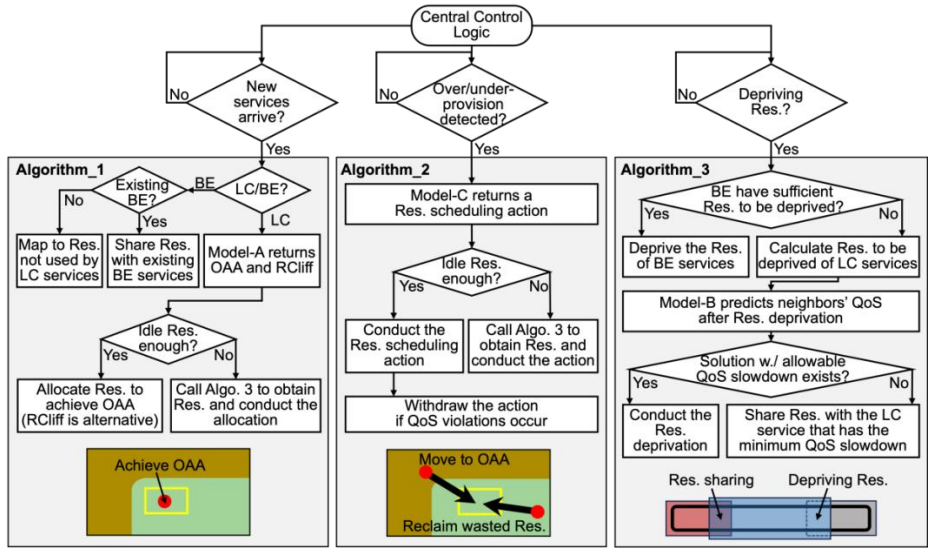

Figure 5. The central control logic of OSML+.

### 3.3 A Multi-model Collaborative Scheduling Framework

**Overview.** Figure 4 shows the overall design of OSML+. OSML+ is a per-node scheduler. OSML+ has a central control framework that coordinates the ML models, manages the data/control flow, and reports the scheduling results. When a new LC service arrives, OSML+ predicts the optimal resource allocation using Model-A. OSML+ enables Model-B to predict the QoS for an LC service when a new resource allocation is conducted, avoiding QoS violations often incurred by the trial-and-error scheduling approaches. When the resource over-/under-provision cases are detected, OSML+ uses Model-C to handle them by adjusting the allocations online. The QoS prediction from Model-B is used as one of Model-C's reward components for faster convergence. OSML+ monitors the status of co-located services for every 100ms by default. It can detect the cases where QoS violations, resource over-provision, or new application launches occur and then enables the related logic. OSML+ performs well with other interval settings and allows configuration flexibility. Figure 5 further shows the framework's overall control logic. More details can be found below.

**Allocating resources for newly coming services.** Algorithm_1 shows how OSML+ uses Model-A in practice. Figure 5 highlights its operations. For a newly coming LC service, OSML+'s central control logic enables Model-A to get the OAA and RCliff. If the current idle resources are insufficient to satisfy the new LC service, OSML+ enables

Model-B to deprive some resources of other services and then allocate them to the new one (refer to Algorithm_3). In terms of BE services (which is not considered in the previous design of OSML [3]), they can achieve a higher throughput when allocated more resources. In the cloud environment, BE services have lower priority than LC services because they do not have strict QoS requirements. Thus, when a BE service arrives, it is mapped to resources not allocated to LC services. If there are existing BE services, OSML+ makes the newly coming BE service share resources with the existing ones. Resource allocations for BE services should not affect the allocations for LC services.

---

Algorithm 1: Allocating resources for newly coming services using Model-A

---

1. **if** the newly coming service is an LC service **then**
2.     Map the newly coming LC service on the idle resources and capture input parameters for Model-A (Table 2);
3.     Forward these parameters as inputs to Model-A;
4.     Model-A outputs: (1) OAA to meet the QoS target; (2) OAA bw (3) RCliff in current environment;
5.     **if** idle resources are sufficient to meet OAA **then**
6.         Allocate resources with a specific solution in OAA (RCliff is alternative);
7.     **else** // Depriving resources from neighboring services using Model-B
8.         Calculate the difference <M,N> between idle resources and the OAA // M CPU cores and N LLC ways are required to meet the LC service's QoS target;
9.         Call Algorithm_3 to obtain M CPU cores and N LLC ways from neighboring services and allocate them to the LC service;
10.     **end**
11. **end**
12. **if** the newly coming service is a BE service **then**
13.     **if** there are existing BE services **then**
14.         Make it share resources with existing BE services;
15.     **else**
16.         Map the BE service to resources not used by LC services.
17.     **end**
18. **end**

---

Algorithm 2: Handling resource under-provision/over-provision using Model-C

---

1. **for** each under-provisioned/over-provisioned LC service **do**
2.     Capture the current status $S_t$ of the LC service and forward $S_t$ to Model-C;
3.     Model-C outputs $A_t$, i.e., the probability distribution of available actions over the action space (Table 5);
4.     Model-C returns the action with the maximum probability;
5.     **if** the action can be satisfied within the current idle resources **then**
6.         OSML+ conducts the action;
7.     **else**
8.         Calculate the resources required to conduct the action, i.e., <M,N> // M CPU cores and N LLC ways are required to conduct the action;
9.         Call Algorithm_3 to obtain M CPU cores and N LLC ways from neighboring services and conduct the action;
10.     **end**
11.     Withdraw the actions if the QoS is violated. // Rollback
11. **end**

---

**Dynamic adjusting for varying loads.** Figure 5 shows the dynamic adjusting of Algorithm_2, in which Model-C is dominant. During the run time, OSML+ monitors each LC service's QoS status for every 100ms. LC services in the cloud environments have changing loads, leading to QoS violations (resource under-provision) and resource waste (resource over-provision). OSML+ detects these cases and enables Algorithm_2 to handle them. An LC service is under-provisioned if its QoS is not satisfied. OSML+ addresses the QoS violations using Model-C. Model-C learns from

---
Algorithm 3: Handling resource deprivation using Model-B
---
1. // OSML+ needs to deprive M CPU cores and N LLC ways from neighboring services
2. **if** BE services have sufficient resources to be deprived **then**
3.     Deprive M CPU cores and N LLC ways from BE services;
4. **else**
5.     Calculate resources required from the LC services, i.e., <M',N'> // M' CPU cores and N' LLC ways are required from neighboring LC services;
6.     **for** each potential neighboring LC service **do**
7.         Create deprivation policies, i.e., {<m,n>|0≤m≤M', 0≤n≤N'};
8.         Use Model-B to infer QoS when the neighbor is deprived of m CPU cores and n LLC ways;
9.     **end**
10.    **if** there is a best-fit solution that involves at most 3 LC services with allowable QoS slowdown **then**
11.        Conduct the resource deprivation;
12.    **else** // Try resource sharing
13.        Share M' CPU cores and N' LLC ways with the neighboring LC service that has the minimum QoS slowdown.
14.    **end**
15. **end**
---

historical experiences to increase the resources for the LC service in these under-provision cases. LC services that meet the QoS targets and have more resources than OAA are over-provisioned (e.g., response latency is lower than QoS target [3,23,26]). OSML+ reclaims the over-provided resources using Model-C and can avoid the RCliff to protect QoS in these cases. Via online learning, Model-C tends to save resources by reclaiming over-provided resources. If QoS violations occur after reclaiming the resources, Algorithm_2 withdraws the action to prevent severe QoS fluctuations. Moreover, Model-C conducts online training. It can correct the resource under/over-provision cases even when the loads dynamically change. As the scheduling time interval is short (100ms), Model-C quickly adjusts the resource allocations for each application within seconds, leading to faster convergence compared with previous approaches.

**Depriving resources for reallocation.** Algorithm_3 deprives resources of existing services in the cases where idle resources are insufficient to satisfy the OAA predicted by Model-A and Model-C's scheduling actions. To this end, Algorithm_3 mainly deprives resources of neighboring BE services. The resources of LC services are deprived seldom. They are only deprived when BE services cannot satisfy the required resources. Algorithm_3 uses Model-B to predict the QoS of neighboring LC services after depriving a specific number of CPU cores and LLC ways, and finds the best-fit solution involving at most 3 LC services with allowable QoS slowdown. If the needs cannot be met involving at most 3 LC services, Algorithm_3 enables resource sharing. The needs are satisfied by sharing resources with the LC service with the minimum QoS slowdown after resource sharing. Resource sharing usually happens between only two applications to minimize the adverse effects. Note that Algorithm_3 might incur resource sharing over the RCliff and thus may bring higher response latency for one or more LC services. OSML+ will report the potential QoS slowdown to the upper scheduler and ask for the decisions. The corresponding actions will not be conducted if the slowdown is not allowed.

### 3.4 Implementation

We design OSML+ to work cooperatively with OS. As the kernel space lacks the support of ML libraries, OSML+ lies in the user space and exchanges information with the OS kernel. OSML+ is implemented using Python and C. It employs Intel CAT technology [29] to control the cache way allocations and supports dynamically adjusting. OSML+ uses Linux's taskset and MBA [30] to allocate specific cores and bandwidth to an LC service. OSML+ captures the online performance parameters using the pqos tool [29] and PMU [30]. The ML models are based on TensorFlow [31] with the version 2.0.4 and can be run on either CPU or GPU. We use Linux with the kernel version of 6.1.

# 4 EVALUATIONS

## 4.1 Methodology

We evaluate OSML+ performance using diverse workloads in the cloud environment, including LC services and throughput-oriented BE services. Details on LC services are in Table 2. BE services used in the evaluation include blackscholes, streamcluster, bodytrack from the Parsec benchmark suite [22]. The metrics involve the QoS of LC services (similar to [3,23], the QoS target of each LC service is the 99th percentile latency of the knee of the latency-RPS curve. Latency higher than the QoS target is a violation.); Effective Machine Utilization (EMU) [23,26] (the max aggregated load of all co-located LC services) – higher is better; and the normalized throughput of the BE services (the throughput is measured by the instructions per clock and is normalized to the solely running case, higher is better). We first evaluate co-location cases of LC and BE services, where LC services run at constant loads. Then, we explore workload churn. We inject LC services with loads from 60% to 100% of their respective max load. To evaluate the generalization of OSML+, we employ some unseen LC services not in Table 2 in our experiments. If the scheduler cannot find an allocation that meets all LC services' QoS after 3 minutes, we signal that the scheduler cannot deliver QoS for that configuration.

We compare OSML+ with the most related approaches in [9,23,3] based on the latest open-source version. The work in [23] employs a "trial-and-error" heuristic approach for resource scheduling, we denote this approach as Heuristic-baseline. It makes incremental adjustments in one resource dimension at a time until QoS is satisfied for all applications. The work in [9] predicts the resource allocation that can meet the QoS and maximize the BE throughput using a Bayesian optimizer. We denote this work as BO-baseline. It conducts various allocation policies and samples each of them; it then feeds the sampling results to a Bayesian optimizer to predict the following scheduling policy.

## 4.2 Experimental Platforms

Table 4 shows details of the three servers used in this work. Server 1 is the platform used in our previous work OSML [3]. Server 2 and Server 3 are new servers used in this work. We deploy OSML+ on all of these three platforms. We have collected training data set on Server 1 in the prior work [3]. The offline models on Server 1 are trained using these traces. For Server 2 and Server 3, we leverage transfer learning to train the offline models based on pre-trained models on Server 1. In this manuscript, we mainly show the experimental results on the new platforms, i.e., Server 2 and Server 3, showing how OSML+ works on a new platform and the generalization ability of our design. We also test OSML+'s model errors and evaluate its performance in handling workload churn across three platforms, showing our design can handle diverse cases across different platforms with various configurations.

## 4.3 OSML+ Effectiveness

We show the effectiveness of OSML+ as follows.

(1) **OSML+ exhibits a shorter scheduling convergence time.** Using ML models, OSML+ achieves OAA quickly and efficiently handles cases with diverse loads. Figure 6-a shows the distributions of the scheduling results of 40 workloads for OSML+, Heuristic-baseline and BO-baseline, respectively. Every dot in Figure 6-a represents a specific workload that contains 3 co-located LC services with different RPS. We launch the applications in turn and use a scheduler to handle QoS violations until all applications meet their QoS targets. The x-axis shows the convergence time; the y-axis denotes the achieved EMU. Generally, OSML+ can achieve the same EMU with a shorter convergence time for a specific load. Figure 6-b shows the violin plots of convergence time for these loads. On average, OSML+ takes 5.8 seconds to converge, while Heuristic-baseline and BO-baseline take 15.0 and 32.1 seconds, respectively. OSML+ converges 2.6× and 5.5× faster than Heuristic-baseline and BO-baseline. OSML+ performs stably – the convergence time ranges from 4.0s (best case) to 13.5s (worst case). By contrast, the convergence time in Heuristic-baseline ranges from 5.1s to 68.9s, and BO-baseline is from 18.5s to 75.5s. OSML+ converges faster mainly because the start point in the scheduling space provided by Model-A is close to OAA. Heuristic-baseline and BO-baseline take a longer convergence time, indicating that they require high scheduling overheads in cloud environments. In cloud, jobs come and go frequently; thus, scheduling occurs frequently, and longer scheduling convergence time often leads to unstable/low QoS.

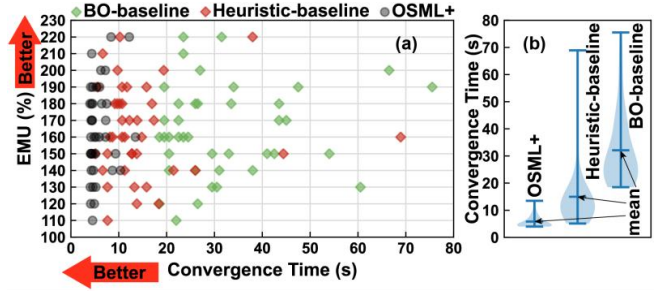

Figure 6: The performance distributions for 40 workloads containing 3 co-located LC services that OSML+, Heuristic-baseline and BO-baseline can all converge.

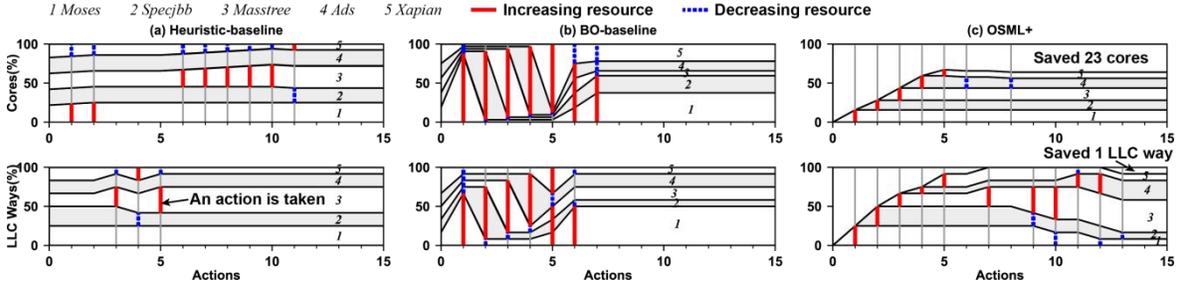

Figure 7: Resource usage comparisons for OSML+, Heuristic-baseline, and BO-baseline.

We further analyze how these schedulers work in detail. Figure 7-a/b/c show the actions used in OSML+, Heuristic-baseline, and BO-baseline's scheduling process for a workload containing five LC services. This case includes Moses, Specjbb, Masstree, Ads, Login with 90%, 60%, 90%, 60%, and 10% of their maximum loads. For this load, Heuristic-baseline, BO-baseline, and OSML+ take 37.0 seconds, 47.6 seconds and 10.6 seconds to converge, respectively. Figure 7 highlights scheduling actions using solid red lines to represent increasing resources and blue dotted lines to denote reducing resources. Figure 7-a shows Heuristic-baseline takes 8 actions for scheduling cores and 3 actions for cache ways. It schedules in a fine-grained way by increasing/decreasing one dimension resource at a time. BO-baseline relies on the sampling points in the scheduling exploration space. Figure 7-b shows BO-baseline repeats sampling until the "expected improvement" in BO-baseline drops below a specific threshold. BO-baseline performs 7 scheduling actions according to its latest open-source version; but it takes the longest convergence time (47.6 seconds). The underlying reason is that BO-baseline's sampling/scheduling doesn't have clear targets. In practice, the improper resource partitions/allocations during sampling lead to the accumulation of requests, and the requests cannot be handled due to resource under-provision. Therefore, it brings a significant increase in response latency. Moreover, due to the early termination of BO-baseline's scheduling process, BO-baseline fails to explore better scheduling solutions that can aid the QoS violations in a more timely manner (e.g., allocating more resources to faster handle the accumulated requests), leading to a long convergence time. Compared with prior schedulers, OSML+ has clear aims and schedules multiple resources simultaneously to achieve them. Figure 7-c shows OSML+ achieves OAA for each LC service with 13 actions, but it has the shortest convergence time – 10.6 seconds. This is because OSML+ has a short scheduling time interval of 100ms by using the predicted QoS as a reward component in Model-C's reward function (Sec. 3.2). It can perform up to ten scheduling actions within a second to achieve OAA quickly. Prior schedulers have longer scheduling time intervals (e.g., 0.5s for Heuristic-baseline and 2s for BO-baseline) because they have to observe the response latency changes after performing a resource scheduling action. Thus, compared with prior schedulers, OSML+ can quickly adjust the resource allocations by performing multiple scheduling actions during the scheduling process.

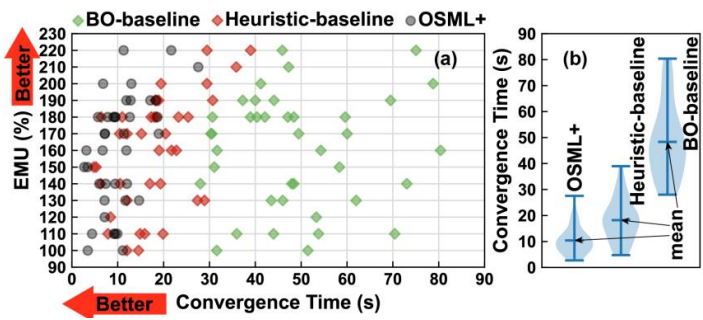
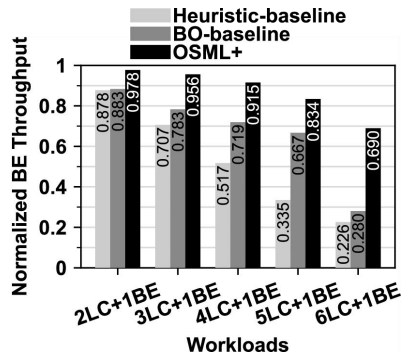

Figure 8: The performance distributions for 40 workloads containing a varying number of LC services and a BE service.

Figure 9. Normalized BE throughput achieved for converged workloads.

OSML+ consumes fewer resources to support the same workloads to meet the QoS targets. As illustrated in Figure 7-a, Heuristic-baseline partitions the LLC ways and cores equally for each LC service at the beginning; once it meets the QoS target (using 11 actions), it stops. Thus, Heuristic-baseline drops the opportunities to explore alternative better solutions (i.e., using fewer cores or cache ways to meet identical QoS targets). Heuristic-baseline allocates all cores and LLC ways finally. BO-baseline also uses all cores and cache ways shown in Figure 7-b. By contrast, OSML+ schedules according to applications' resource requirements instead of using all resources. Figure 7-c shows that using Model-A's OAA prediction and Model-C's online scheduling, OSML+ achieves each LC service's OAA after 13 actions. OSML+ detects and reclaims over-provided resources using Model-C. For example, actions 6 and 8 in Figure 7-c reclaim two cores from Ads, action 13 reclaims one LLC way from Specjbb. Finally, OSML saves 23 cores and 1 LLC way. As OSML is designed for LC services that are executed for a long period, saving resources means saving budgets for cloud providers.

**(2) OSML+ can converge faster and have scheduling results for more workloads that have LC and BE services.** The cost-efficiency policy drives cloud providers to co-locate as many applications as possible on a server, including a varying number of LC services and BE services. Even in these complicated cases, OSML+ can converge faster and support more workloads with LC services compared with previous schedulers. In terms of BE services, they can achieve a higher throughput when allocated more resources. OSML+ allocates more resources to BE services by reclaiming over-provided resources of LC services, improving the BE throughput. Figure 8-a shows the distributions of scheduling results of 40 workloads in these complicated co-location cases. The workloads include cases where two, three, four, five, or six LC services are co-located with one BE service. The BE service can be either blackscholes, streamcluster, or bodytrack [22]. Figure 8-b shows the violin plots of convergence time for these workloads. On average, OSML+ takes 10.4 seconds to converge, while Heuristic-baseline and BO-baseline take 18.2 and 48.4 seconds, respectively. The convergence time for OSML+ ranges from 2.7s (best case) to 27.6s (worst case). By contrast, the convergence time in Heuristic-baseline ranges from 4.8s to 39.0s, and BO-baseline is from 28.0s to 80.4s. OSML+ takes a shorter time for convergence in these complicated co-location cases, showing that OSML+ is generalizable and performs stable in various cases.

As OSML+'s scheduling is fast, it supports more workloads. We evaluate cases where a workload has two, three, four, five, and six LC services co-located with one BE service, respectively. The BE service can be any one of blackscholes, streamcluster, or bodytrack [22]. For each case, we schedule 20 workloads using OSML+, Heuristic-baseline and BO-baseline. OSML+ can converge for all workloads when there are two, three, four and five LC services in the workload, and only fails to converge for one workload when there are six LC services in the workload. By contrast, for workloads where there are two, three, four, five, and six LC services co-located with one BE service, Heuristic-baseline fails to converge for 0, 1, 4, 6, and 8 workloads, respectively; BO-baseline fails to converge for 0, 2, 4, 5, and 7 workloads, respectively. OSML+ outperforms the previous schedulers mainly because it can quickly achieve OAA in the large scheduling exploration space using ML models. When there are a larger number of co-located services in the workload, resource scheduling becomes complicated as the co-located services contend for the interactive resources. Heuristic-baseline and BO-baseline cannot achieve OAA directly in the large scheduling exploration space. They have to sample in a fine-grain way, inevitably incurring QoS fluctuations during the scheduling process. By contrast, OSML+ achieves OAA using offline ML models and can effectively avoid QoS violations. Thus, it can converge for more workloads. Moreover, OSML+ can achieve higher throughput for BE services by reclaiming over-provided resources using Model-C. Figure 9 shows the achieved throughput for BE services for the converged workloads with a varying number of LC services. The BE throughput is normalized to the solely running cases. For workloads where there are two, three, four, five, and six LC services co-located with one BE service, OSML+ achieves an average BE throughput $1.1\times$, $1.4\times$, $1.8\times$, $2.5\times$, $3.1\times$ higher than Heuristic-baseline, respectively; and achieves $1.1\times$, $1.2\times$, $1.3\times$, $1.3\times$, $2.5\times$ higher than BO-baseline, respectively.

**(3) The Comparisons between OSML+ and OSML [3].** OSML+ is a derivative version of the OSML project. OSML+ is an extension of OSML. We compare OSML+ with the prior study OSML [3]. OSML+ converges faster for LC services. For workloads containing three co-located LC services, OSML and OSML+ take an average of 20.9 seconds and 5.8 seconds to converge, respectively. OSML+ converges $3.6\times$ faster than OSML mainly because it has a short scheduling time interval of 100ms. OSML schedules every 1 second because it has to wait and observe the response latency changes to obtain the feedback for a resource scheduling action [3]. By contrast, OSML+ works well with the 100 ms interval because it uses the predicted QoS as a reward component in Model-C (Sec. 3.2). OSML+ can obtain the predicted feedback instantly using Model-B without waiting for the response latency changes. With the 100ms scheduling time interval, OSML+ can conduct more scheduling actions during the same scheduling phase to achieve OAA faster. Moreover, OSML+ supports complicated co-location cases, including LC and BE services. By reclaiming over-provided resources and allocating them to BE services, OSML+ achieves an average BE throughput up to $3.1\times$ and $2.5\times$ higher than Heuristic-baseline and BO-baseline, respectively. By contrast, BE services are not considered in OSML [3]. Moreover, regarding memory consumption, OSML+ uses about 164.6 MB and OSML uses around 82.5 MB. OSML+ has a larger memory footprint, as OSML+'s memory pool in Model-C is larger than OSML's memory pool in Model-C. In terms of computing overheads, OSML+'s CPU usage is 1.45% of a single CPU core on Server 2 in Table 4. For OSML, it is 0.96%. OSML+ has higher computing overheads mainly because it has a shorter scheduling time interval of 100 ms (the scheduling time interval for OSML is 1 s). Both OSML+ and OSML are lightweight. They have marginal impacts on the applications running on the platforms.

### 4.4 Performance for Workload Churn

We evaluate how OSML+ behaves with dynamically changing loads on the three platforms in Table 4 using the same workload. Server 1 in Table 4 is a ten-year-ago platform primarily used in the previous work [3]. Server 2 and Server 3 are new platforms. We focus on the experimental results on Server 2 (Figure 10-2). As illustrated in Figure 10-2 (a), in the beginning, five LC services including Img-dnn, Login, Ads, Specjbb, and Xapian with 60%, 60%, 60%, 30% and 30% of their max load arrive. OSML+ quickly meets the QoS demands for all LC services by using Model-A's OAA prediction and Model-C's online scheduling. Figure 10-2 (b) shows that all LC services have sufficient resources to meet their QoS targets. Heuristic-baseline aids the QoS violation of Img-dnn step by step, i.e., adjusting one dimension resource at a time until the QoS is satisfied. It allocates resources to Img-dnn by depriving resources of co-located services, which takes 8 seconds to aid the QoS violations. As illustrated in Figure 10-2 (c), Img-dnn has high response latency in the beginning but the QoS violation is corrected at 12s. BO-baseline's scheduling relies on sampling and Bayesian optimizer. It starts scheduling at time point 0, where all the five LC services arrive. BO-baseline conducts 10 sampling steps in 20 seconds, but cannot obtain a scheduling solution that satisfies these services' QoS targets. In Figure 10-2 (d), Img-dnn and Specjbb have high latency. During the same scheduling phase (e.g., the time point 0 to 20), we can observe that OSML+ has the lowest overall normalized latency in Figure 10-2 (b). Using ML models' OAA prediction, OSML+ quickly meets the QoS targets.

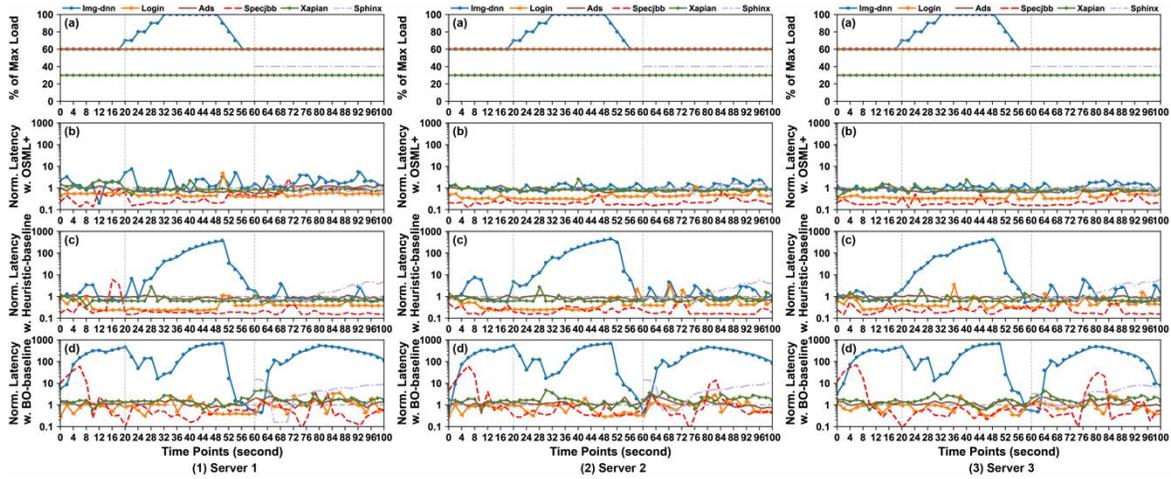

Figure 10. OSML+'s performance in reality with varying loads across platforms in Table 4. Each LC service's response latency is normalized to the solely running case. Figure 10-1/2/3 denote the performance on Server 1/2/3, respectively.

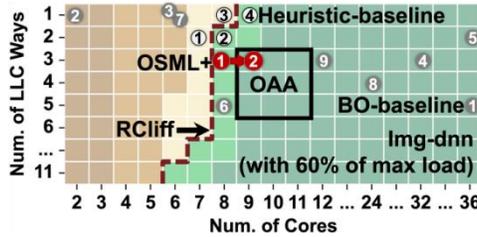

Figure 11. The scheduling traces in the scheduling space for all schedulers on Server 2 from time point 60 to 100 in Figure 10. Each circle denotes a specific scheduling policy conducted by a specific scheduler. The number in a circle denotes the sequence of these scheduling actions during the scheduling phase.

From 20 to 32, we increase the load for Img-dnn as illustrated in Figure 10-2 (a). OSML+ meets Img-dnn's changing demands by using Model-C. As illustrated in Figure 10-2 (b), all LC services have low response latency. Heuristic-baseline does not reflect quickly for this change. It adjusts resources step by step and cannot efficiently schedule resources to handle QoS violations in a timely manner. Thus, as illustrated in Figure 10-2 (c), the QoS violation is not aided until 58s, when Img-dnn's load decreases. For BO-baseline, it has to sample each time when the load changes. But during the sampling, a specific service might not have sufficient resources to handle the requests; thus the requests are accumulated, leading to QoS fluctuations/violations during the scheduling. Figure 10-2 (d) shows that Img-dnn cannot meet its QoS targets until 58 seconds.

At time point 60, Sphinx with 40% of its max load arrives as illustrated in Figure 10-2 (a). As mentioned before, OSML+ saves resources and thus it can serve more workloads. OSML+ allocates saved resources to Sphinx without depriving resources of other LC services. It meets Sphinx's demands without causing QoS violations for neighboring LC services. Figure 10-2 (b) shows that all LC services have low response latency. Even there are QoS violations for Img-dnn, they can be quickly corrected by using Model-C. Heuristic-baseline allocates resources to Sphinx by depriving resources of neighboring LC services, incurring QoS fluctuations of Img-dnn, as shown in Figure 10-2 (c). BO-baseline samples several scheduling policies in the scheduling space, but does not converge and thus incurs QoS fluctuations. As illustrated in Figure 10-2 (c), Img-dnn and Sphinx fail to meet their QoS targets, and Sphinx's response latency keeps increasing due to resource under-provision. Figure 11 highlights the scheduling actions for Img-dnn from 60 to 100. During this phase, Heuristic-baseline cannot avoid the RCliff and thus incurs QoS fluctuations for Img-dnn. BO-baseline samples nine resource allocation policies for Img-dnn in the scheduling space. At 82s, it finds an allocation policy that can correct the QoS violation, i.e., Img-dnn's response latency begins to decrease at 82s. But Img-dnn's QoS is not met at 100s due to the accumulated requests during the scheduling process. By contrast, OSML+ achieves Img-dnn's OAA using only one scheduling action. All LC services have low response latency during this phase.

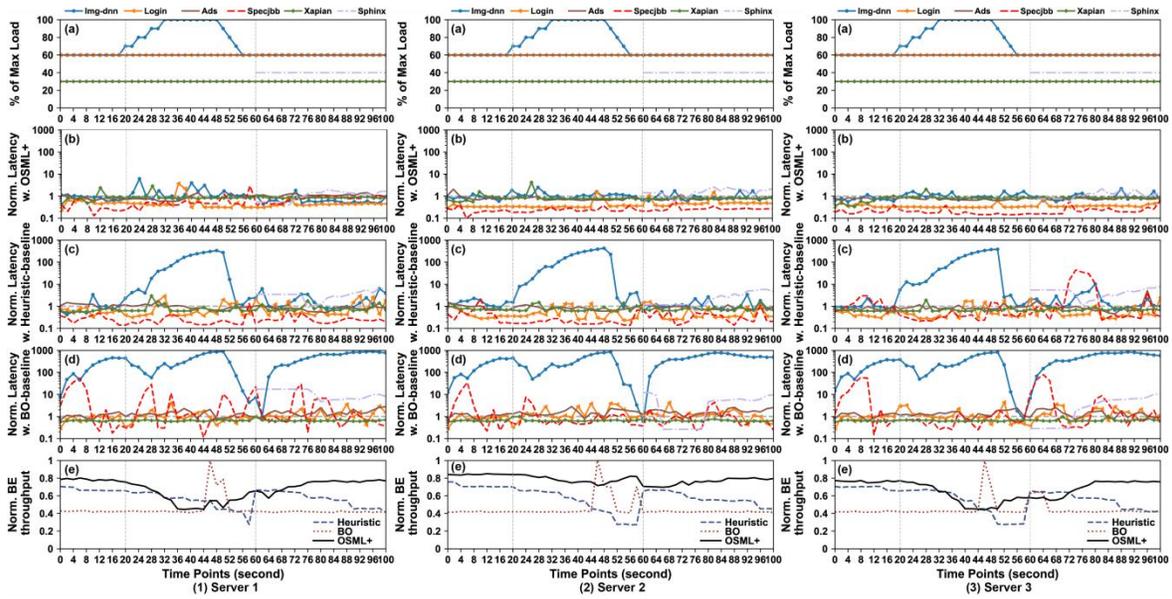

Figure 12. OSML+'s performance for a workload with both LC and BE services. The BE service's throughput is normalized to the solely running case. Figure 12-1/2 and 3 denote the performance on Server 1/2 and 3, respectively.

Experiments on Server 1 and Server 3 exhibit similar results. We observe LC services on Server 2 and Server 3 achieve lower response latency when scheduled using OSML+. For example, when the Img-dnn′s load increases at 20s, we observe higher latency for Img-dnn on Server 1 than on Server 2 and Server 3, even though the QoS violation is quickly corrected using Model-C within 4 seconds. From 60 to 100, Img-dnn also exhibits higher response latency than on Server 2 and Server 3. This indicates that models trained on new platforms using transfer learning are more accurate and can better handle resource scheduling for co-located services. Transfer learning makes the ML models generalizable across new platforms with the minimum data collection overhead.

We evaluate how OSML+ behaves when the workload has both LC and BE services. We conduct the experiments on three platforms in Table 4 and focus on experimental results on the recent Server 2. We use the same LC services as Figure 10 and launch blackscholes as a BE service at time point 0. The BE service can achieve a higher throughput when allocated with more resources. OSML+ meets each LC service's QoS target using the minimum resources and can reclaim the over-provided resources of LC services. Therefore, OSML+ can provide more resources to the BE service, improving the BE throughput without violating LC services' QoS. In the beginning, OSML+ meets the LC services' demands using Model-A and allocates the saved resources to the BE service. As illustrated in Figure 12-2 (e), OSML+ achieves a higher BE throughput than Heuristic-baseline and BO-baseline. In Figure 12-2 (c), Heuristic-baseline deprives resources of the BE service to aid Img-dnn's QoS violation. Thus, the BE throughput decreases. In Figure 12-2 (d), BO-baseline samples in the scheduling space to aid Img-dnn and Specjbb's QoS violations. It does not improve throughput for the BE service. From 20 to 32, OSML+ deprives resources of the BE service to satisfy Img-dnn's increasing demands. Therefore, the BE throughput of OSML+ decreases during this phase as shown in Figure 12-2 (e). From 50 to 58, Img-dnn's load decreases. OSML+ detects that the resources are over-provisioned. Thus, it reclaims the over-provided resources and allocates them to the BE service, improving the BE service's throughput. By contrast, from 20 to 60, Heuristic-baseline deprives resources of the BE service to aid Img-dnn's QoS violations, reducing the BE throughput. BO-baseline samples several scheduling policies, but cannot meet the LC services' QoS targets. From 60 to 100, OSML+ can keep a high BE throughput by reclaiming over-provided resources using Model-C. Heuristic-baseline deprives resources of the BE service and allocates them to Img-dnn and Sphinx, reducing the BE throughput. BO-baseline can not find a solution that satisfies the QoS during this phase. In general, OSML+ exhibits the highest overall BE throughput during the scheduling process. Experiments on Server 1 and Server 3 exhibit similar results, showing that OSML+ works stably across various platforms. More discussions on generalization refer to the following section. Moreover, Figure 12-1/2/3 (e) shows that OSML+ achieves a higher BE throughput on Server 2 than on Server 1 and Server 3. This is because Server 2 has more cores; thus, OSML+ can achieve a higher BE throughput.

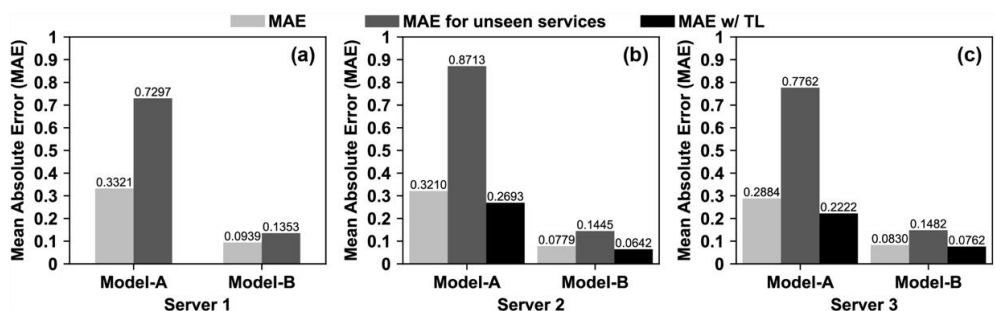

Figure 13. Mean Absolute Error (MAE) on the test set across platforms (Table 4). TL is used on Server 2 and 3, so their sub-figures has the bars for MAE w/ TL.

### 4.5 Generalizable ML Models

**Offline models.** We show that OSML+'s offline models are accurate and generalizable by evaluating their Mean Absolute Error (MAE) [3] in the following cases: (1) models are trained without transfer learning, (2) models are tested using unseen cloud services that are not involved during the training process, and (3) models are trained on new platforms using transfer learning. The results are in Figure 13. For case (1), we have collected extensive data sets for training Model-A and Model-B on Sever 1 [3]. By contrast, on Server 2 and Server 3, we only collected data for the cases where the number of launched threads equals to the number of CPU cores on each platform, which only takes hours. We use "hold-out validation" for evaluating the offline models [3], i.e., 70% of the whole data set is used as the training set, and the remaining 30% is used as a test set. We train the offline models using the training set and evaluate their MAE using the test set on each platform. The results are shown as "MAE" in Figure 13. For case (2), we evaluate the models' MAE on a test set containing traces of unseen LC services that are not in Table 2 (including MySQL [41], Redis [42], and Silo [37]). The results are shown as "MAE for unseen services" in Figure 13. For case (3), we train the models on two new platforms (Server 2 and Server 3) using transfer learning based on the pre-trained models on Server 1. We copy the network parameters of offline models on Server 1 to the models on the new platforms, and train them using the training set from the new platforms. We evaluate the models' MAE on the test set on each platform and show the results as "MAE w/ TL" in Figure 13. The experimental results show that OSML+'s offline models are generalizable across new platforms by using transfer learning. As illustrated in Figures 12-b and c, compared with training models without transfer learning, using transfer learning can achieve an average of 19.5% and 12.9% reduction in MAE for Model-A and Model-B, respectively, on the two new platforms. Moreover, OSML+'s offline models are generalizable for unseen services. For Model-A, the MAE for unseen services on the three platforms are 0.7297, 0.8713, and 0.7762, respectively. Even though the offline models have higher errors for unseen services, the error for the predicted OAA is less than one core or one LLC way. This indicates that Model-A can still predict (near) optimal solutions in practice. For Model-B, the MAE for unseen services on the three platforms is 0.1353, 0.1445, and 0.1482, respectively. This indicates that the average errors in Model-B's predicted QoS for unseen services on the three platforms are 13.53%, 14.45% and 14.82%, respectively. In practice, the prediction is accurate enough to detect QoS violations that often have $10\times \sim 100\times$ of QoS degradation.

**Online learning Model-C.** We show that Model-C can achieve faster convergence by using reinforcement learning in Figure 14. We first train Model-C on Server 2 on workloads containing three LC services co-located with one BE service. The blue curve (lower curve) in Figure 14 shows Model-C's episode reward during the online training process. Each episode has 200 training steps. The episode reward is the sum of the reward value in each training step of an episode. A higher episode reward indicates that Model-C can achieve better QoS for LC services and save more resources during this episode. Model-C converges after 300 episodes, where the episode reward stabilizes and consistently remains high. When using transfer learning, Model-C can converge much faster. We train Model-C on Server 2 using transfer learning. We first copy the network parameters of Server 1's pre-trained Model-C to the networks on Server 2, and then conduct online training. The orange curve (higher curve) in Figure 14 shows that Model-C can quickly converge in 150 episodes, which takes around one hour in practice. Moreover, Model-C improves the episode reward by 20.7% on convergence. This indicates that Model-C can be easily deployed on the new platform using transfer learning with hours of online training.

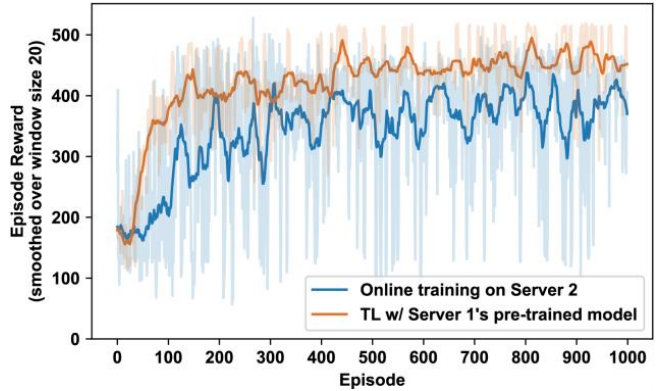

Figure 14. The episode reward of Model-C. Each episode has 200 training steps.

## 5 THE EXPERIENCES ON BUILDING AN INTELLIGENT SYSTEM VIA THE STUDIES OF OSML/OSML+

In this section, we summarize our experiences making OS intelligent based on our practice building OSML+.

(1) The first one is to *break down the big/complicated task and use small, lightweight models to handle sub-tasks*, respectively. OSML/OSML+ employs three lightweight ML models to handle sub-tasks during the resource scheduling. Each model has its respective duty. As the models are lightweight and their functions are clearly defined, it is easy to locate and debug the problems. So the system can be robust.

(2) If more than one ML model is used in the design, having a framework as the central point to coordinate models and work with other OS components is necessary. In our design, the framework has clearly defined interfaces for managing and coordinating ML models. New models can be added to the framework for new tasks, and new models for different optimization goals can replace existing models. Thus, the framework is generalizable across various optimization tasks in OS design. For new platforms, transfer learning can be used to train models on the new platforms and then replace the old models. To sum up, *the takeaway is having a good design on the central framework, which is flexible, robust, generalizable, and clean in control logic.*

(3) Training data is essential to the performance of ML models, especially in the cases where the system will be generalized onto diverse platforms. *Open source is a good way*. In OSML/OSML+, We have collected the performance traces for widely deployed LC services, covering 62,720,264 resource allocation cases that contain around 2 billion samples. These data have a rich set of information, e.g., the RCliffs for multiple resources, the interactions between workload features, and the mainstream architectures. ML models can be trained and generalized with these data and then used on new platforms with low-overhead transfer learning. The date set will be open-sourced. People can study the data set and train their models without a long data collection period. Moreover, *OSML/OSML+ is a long-term project open to the community; we continue adding new traces collected from new applications and servers to the data set to enhance models' performance for new cases.*

# 6 CONCLUSION

We present OSML+, an ML-based resource scheduler for co-located LC services. We learn that straightforwardly using a simple ML model might not handle all of the processes during the scheduling. Therefore, using multiple ML models cooperatively in a pipe-lined way can be an ideal approach. More importantly, we advocate a new solution, i.e., leveraging ML to enhance resource scheduling, which could have immense potential for OS design. We think OS can be more intelligent to meet emerging applications. In a world where co-location and sharing are fundamental realities, our solution, i.e., OSML/OSML+, should grow in importance and benefit our community. Moreover, we also think that some large model technologies can be used in future studies for OSML. OSML+ and OSML can be accessed via the link - https://github.com/Sys-Inventor-Lab/AI4System-OSML.

# ACKNOWLEDGMENT

We thank the reviewers, AE and EIC, for their valuable comments. Xinglei Dou is a student member in Sys-Inventor Lab led by Lei Liu (PI). This project is supported by Beijing Natural Science Foundation under Grant No. L243001, the Key-Area Research and Development Program of Guangdong under Grant No. 2021B0101310002, and the NSF of China under Grant No. 62072432. Lei Liu is the corresponding author and co-first author.# REFERENCES

[1]   Haoran Qiu, Subho S. Banerjee, Saurabh Jha, Zbigniew T. Kalbarczyk, and Ravishankar K. Iyer. "FIRM: An Intelligent Fine-grained Resource Management Framework for SLO-Oriented Microservices," in OSDI, 2020.

[2]   Mingzhe Hao, Levent Toksoz, Nanqinqin Li, Edward Edberg Halim, Henry Hoffmann, and Haryadi S. Gunawi. "LinnOS: Predictability on Unpredictable Flash Storage with a Light Neural Network," in OSDI, 2020.

[3]   Lei Liu, Xinglei Dou, and Yuetao Chen, "Intelligent Resource Scheduling for Co-located Latency-critical Services: A Multi-Model Collaborative Learning Approach," in FAST, 2023.

[4]   Jian Tan, Tieying Zhang, Feifei Li, Jie Chen, Qixing Zheng, Ping Zhang, Honglin Qiao, Yue Shi, Wei Cao, and Rui Zhang, "iBTune: Individualized Buffer Tuning for Large-scale Cloud Databases," in VLDB Endowment, 2019.

[5]   Romil Bhardwaj, Kirthevasan Kandasamy, Asim Biswal, Wenshuo Guo, Benjamin Hindman, Joseph Gonzalez, Michael I. Jordan, and Ion Stoica, "Cilantro: Performance-Aware Resource Allocation for General Objectives via Online Feedback," in OSDI, 2023.

[6]   Haoran Qiu, Weichao Mao, Chen Wang, Hubertus Franke, Alaa Youssef, Zbigniew T. Kalbarczyk, Tamer Basar, and Ravishankar K. Iyer, "AWARE: Automate Workload Autoscaling with Reinforcement Learning in Production Cloud Systems," in ATC, 2023.

[7]   Rajiv Nishtala, Vinicius Petrucci, Paul M. Carpenter, Magnus Själander, "Twig: Multi-Agent Task Management for Colocated Latency-Critical23